\documentclass[fleqn,usenatbib]{mnras}

\usepackage[T1]{fontenc}
\usepackage{ae,aecompl}

\usepackage{graphicx}	
\usepackage{amsmath}	
\usepackage{amssymb}	

\newcommand{\etal} {et~al.~}
\def \spose#1{\hbox  to 0pt{#1\hss}}  
\newcommand{\lta} {\mathrel{\spose{\lower 3pt\hbox{$\sim$}}\raise  2.0pt\hbox{$<$}}}
\newcommand{\gta} {\mathrel{\spose{\lower  3pt\hbox{$\sim$}}\raise 2.0pt\hbox{$>$}}}

\newcommand{\Msun} {\ifmmode \,{\rm M}_{\odot} \else ${\rm M}_{\odot}$ \fi}
\newcommand{\Lsun} {\ifmmode \,{\rm L}_{\odot} \else ${\rm L}_{\odot}$ \fi}
\newcommand{\LV} {\ifmmode L_{\rm V} \else $L_{\rm V}$ \fi}
\newcommand{\Mstar} {\ifmmode M_{\rm star} \else $M_{\rm star}$ \fi}
\newcommand{\Mhalo} {\ifmmode M_{\rm halo} \else $M_{\rm halo}$ \fi}
\newcommand{\LCDM} {\ifmmode \Lambda \rm CDM \else $\Lambda$CDM \fi}
\newcommand{\kms} {\ifmmode \,{\rm km\,s^{-1}} \else km\, s$^{-1}$ \fi} 

\title[Too big doesn't fail]{NIHAO V: Too big doesn't fail -- reconciling the conflict between $\Lambda$CDM predictions and the circular velocities of nearby field galaxies}

\author[Aaron A. Dutton et al.]{Aaron A. Dutton,$^{1,2}$\thanks{E-mail: dutton@nyu.edu}
Andrea V. Macci\`o,$^{1,2}$
Jonas Frings,$^{2,3}$
Liang Wang,$^{2,4}$
\newauthor{Gregory S. Stinson,$^2$ Camilla Penzo,$^2$ Xi Kang,$^4$}\\
$^1$New York University Abu Dhabi, PO Box 129188, Abu Dhabi, UAE\\
$^2$Max-Planck-Institut f\"ur Astronomie, K\"onigstuhl 17, 69117 Heidelberg, Germany\\
$^3$Astronomisches Recheninstitut, Zentrum f\"ur Astronomie der Universit\"at
Heidelberg, Philosophenweg 12, 69120 Heidelberg, Germany\\
$^4$Purple Mountain Observatory, 2 West Beijing Road, Nanjing 210008, China\\
}

\date{Accepted 2015 November 27. Received 2015 November 12; in original form 2015 October 17} 

\pubyear{2015}

\begin{document}
\label{firstpage}
\pagerange{\pageref{firstpage}--\pageref{lastpage}}
\maketitle

\begin{abstract}
  We compare the half-light circular velocities, $V_{1/2}$, of dwarf
  galaxies in the Local Group to the predicted circular velocity
  curves of galaxies in the NIHAO suite of \LCDM simulations. We use a
  subset of 34 simulations in which the central galaxy has a stellar
  luminosity in the range $0.5\times 10^{5} < \LV/\Lsun <
  2\times 10^{8}$.  The NIHAO galaxy simulations reproduce the
  relation between stellar mass and halo mass from abundance matching,
  as well as the observed half-light size vs luminosity relation. The
  corresponding dissipationless simulations over-predict the
  $V_{1/2}$, recovering the problem known as too big to fail (TBTF).
  By contrast, the NIHAO simulations have expanded dark matter haloes,
  and provide an excellent match to the distribution of $V_{1/2}$ for
  galaxies with $\LV \gta 2\times 10^{6}\Lsun$.  For lower
  luminosities our simulations predict very little halo response, and
  tend to over predict the observed circular velocities.  In the
  context of $\LCDM$, this could signal the increased  stochasticity
  of star formation in haloes below $\Mhalo \sim 10^{10}\Msun$, or the
  role of environmental effects. Thus, haloes that are ``too big to
  fail'', do not fail $\LCDM$, but haloes that are ``too small to
  pass'' (the galaxy formation threshold) provide a future test of
  $\LCDM$.
\end{abstract}




\section{Introduction}

The Dark Energy plus Cold Dark Matter ($\LCDM$) model provides an
extremely successful cosmological framework for understanding the
large ($>$ Mpc) scale structure of the universe and its evolution with
time.  On small (kpc) scales the \LCDM model has faced challenges
related to the number density and structure of dark matter haloes. At
face value \LCDM predicts too many low mass haloes
\citep{Moore99,Klypin99} and too much mass on scales near galaxy
half-light radii \citep{deBlok01}.  While these ``missing satellite''
and ``cusp-core'' problems may signal the need for alternatives to
cold dark matter \citep[e.g.,][]{Vogelsberger14,Maccio15}, there are
plausible solutions related to the baryonic physics of galaxy
formation.

Recently, \citet{BK11,BK12}, introduced a related problem.  Using
dissipationless \LCDM simulations they found that the 10 most massive
sub-haloes in simulated Milky Way mass haloes have circular velocities
a factor of $\sim 1.5$ higher than that observed  at the half-light
radii of the MW satellites.  This is often referred to as the too big
to fail (TBTF) problem because the haloes are too big ($V_{\rm max} >
30\kms$) for the effects of the cosmic UV background to suppress gas
cooling and thus prevent star formation \citep{Bullock00}. Thus each
halo must host a visible galaxy.  While possibly related to the
missing satellites problem, in that the largest subhaloes may not have
been found, TBTF is a distinct problem related to the internal
structure of subhaloes, and hence to the cusp-core problem, rather
than strictly to their abundances.

A somewhat trivial solution to the TBTF problem is to reduce the mass
of the Milky Way, with proportionally fewer massive subhaloes
\citep{Vera-Ciro13}. However, a lower Milky Way halo mass
significantly reduces the likelihood of finding two subhaloes as
massive as the LMC and SMC \citep{Kennedy14}. A comprehensive
statistical analysis of the sub-halo population indeed concludes that
there are too many massive CDM sub-haloes \citep{Jiang15}.

On the galaxy formation side there are processes that can solve the
TBTF problem. Gas outflows or bulk motions driven by feedback from
stars can cause halo expansion
\citep[e.g.,][]{Navarro96,Mashchenko06,Pontzen12,Ogiya14}. A number of
studies using fully cosmological galaxy formation simulations have
indeed found halo expansion in isolated dwarf galaxies
\citep[e.g.,][]{Mashchenko08,Governato10,DiCintio14b,Madau14,Onorbe15,Tollet15,
  Trujillo15}. In addition  \citet{Zolotov12} and \citet{Brooks14}
studied the satellite population in two Milky Way mass
simulations. The combination of feedback before infall and tidal
stripping after infall resulted in reduced circular velocities, at the
scale of $1 \rm kpc$, of the magnitude required to resolve the TBTF
problem.

Recently it has been shown that field galaxies follow the same trends
between velocity dispersion and half-light radius as satellite
galaxies \citep{Kirby14}, and thus the over-prediction of galaxy
circular velocities persists in the field \citep{GK14b}, see also
\citet{Papa15}. This Field TBTF problem is a cleaner test of $\LCDM$,
as the mass of the Milky Way and environmental processes are not
plausible solutions.  Within the \LCDM framework halo expansion driven
by feedback from stars and supernova is  the only solution. If this
solution fails, then an alternative to Cold Dark Matter is required.

A key question for the feedback solution is whether there is enough
energy available in low mass galaxies to drive sufficient outflows.
Idealized simulations and energy arguments have been used to conclude
that the answer is no \citep{BK12, Penarrubia12, GK13}. However,
subsequent studies have challenged these conclusions, arguing that
there is in fact sufficient energy available to cause halo expansion
on mass scales relevant to the TBTF problem \citep{Madau14,
  Maxwell15,Chan15}.

\begin{figure}
  \centerline{
    \includegraphics[width=0.48\textwidth]{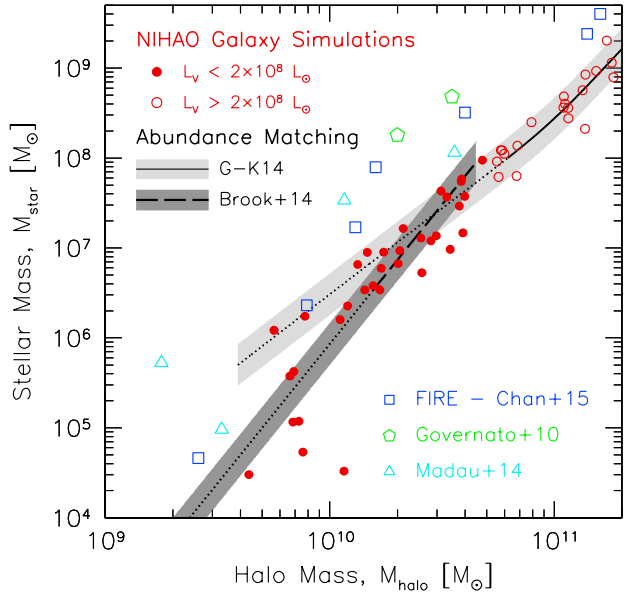}
    }
    \caption{ Stellar mass vs halo mass for NIHAO galaxy simulations
      (red circles) with luminosities $\LV < 2\times 10^{8}\Lsun$
      (filled) and $\LV > 2\times 10^8\Lsun$ (open). NIHAO galaxies
      compare well  to the latest halo abundance matching results
      \citep{Brook14,GK14a}, while other cosmological dwarf galaxy
      simulations tend to over-predict the stellar masses.}
    \label{fig:msmh}
\end{figure}

Thus, on an individual basis the field TBTF problem can be solved, but
what about for the full population of \LCDM haloes, with realistic
galaxy masses and sizes? In this letter we answer this question with a
subset of the NIHAO (Numerical Investigations of Hundred Astrophysical
Objects)\footnote{Nihao is the Chinese word for {\it hello}} galaxy
formation simulations \citep{Wang15}. Reproducing the stellar masses
is critical, as  if star formation is too efficient one will likely
overpredict the amount of expansion, and draw erroneous
conclusions. Previous simulations tend to over-predict the stellar
masses by up for a factor of $\sim 10$ (see Fig.~\ref{fig:msmh}).

\begin{figure*}
\centerline{
  \includegraphics[width=0.49\textwidth]{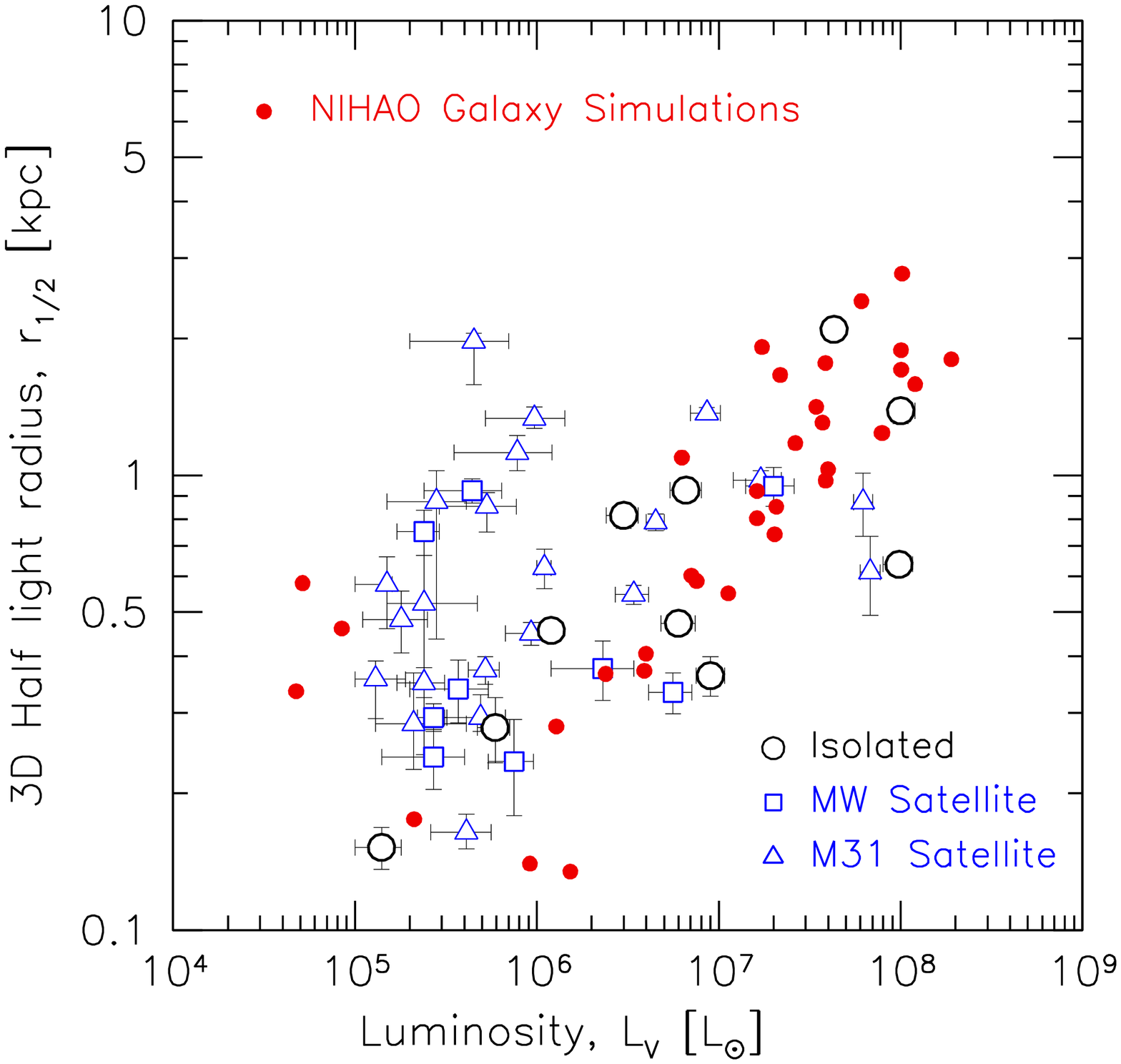}
 	\includegraphics[width=0.49\textwidth]{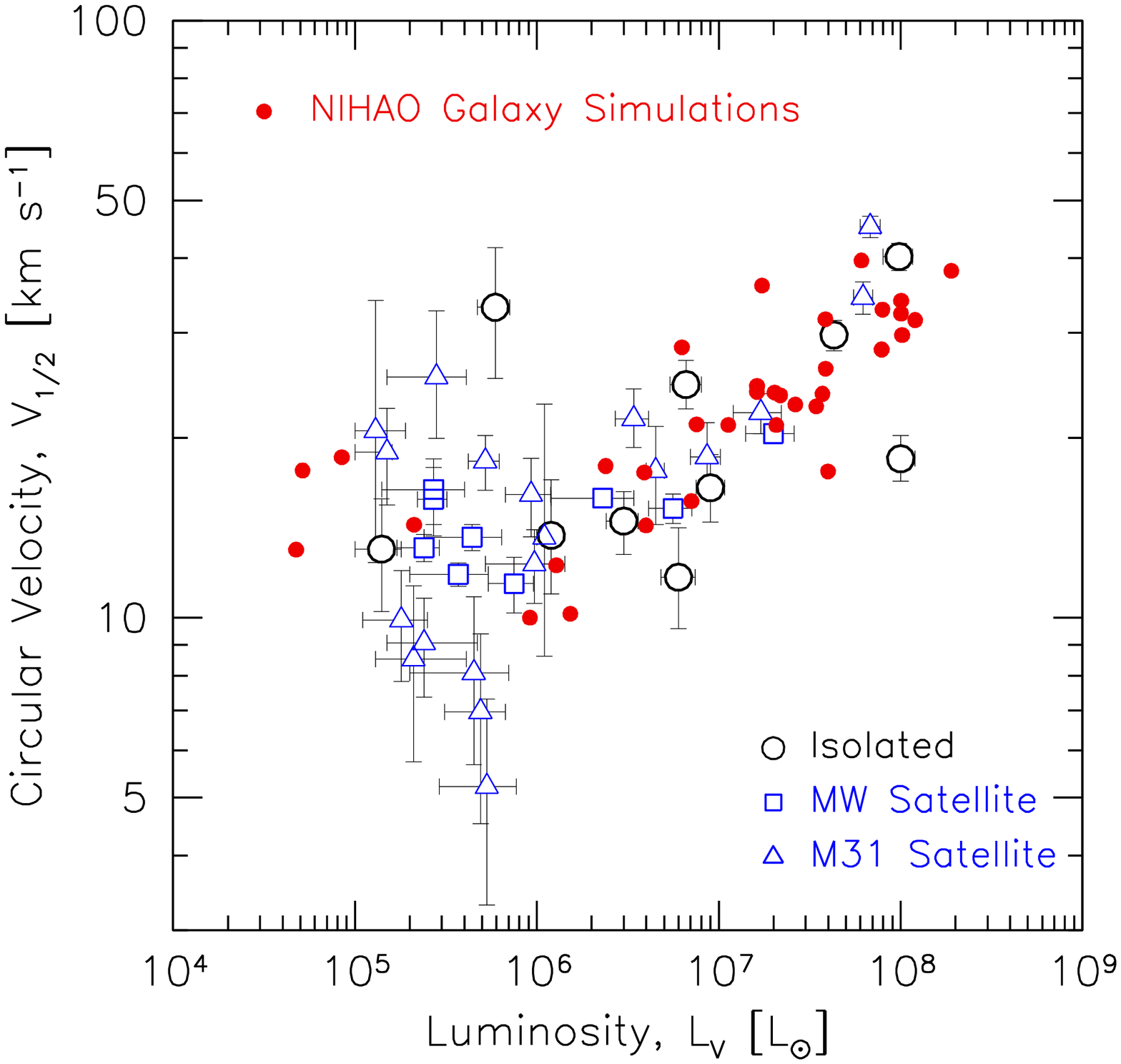} 
}
\caption{Size vs luminosity (left) and velocity vs luminosity (right)
  relations for NIHAO galaxy simulations (red symbols) compared to
  observed galaxies in the local group compiled by \citet{Kirby14}
  split into isolated (black) and satellites of the Milky Way (blue
  squares) and M31 (blue triangles). }
    \label{fig:vlr}
\end{figure*}

\section{Cosmological Simulations}

The NIHAO simulation suite is an unbiased sample of hydrodynamical
cosmological zoom-in simulations of isolated haloes of present day
masses between $\Mhalo \simeq 4\times 10^9 \Msun$ and $\Mhalo \simeq
3\times 10^{12} \Msun$. Haloes are selected independent of mass
accretion history and halo structure.  The resolution was chosen to
resolve the mass profile down to 1\% of the virial radius, with dark
matter force softening of $\epsilon_{\rm dm} = 116-311$pc, and
  hydro particle softenings of $\epsilon_{\rm gas}=50-133$ pc. The
haloes thus typically have a million dark matter particles inside the
virial radius. Two simulations were run for each initial condition: a
dark matter only using {\sc pkdgrav} \citep{Stadel01}; and galaxy
formation using {\sc gasoline} \citep{Wadsley04,Keller14} with the
MaGICC star formation and feedback model \citep{Stinson13}. The free
parameters in the feedback scheme were chosen so that a Milky Way mass
galaxy fits the stellar mass - halo mass relation at $z=0$.  We refer
the reader to \citet{Wang15} for more details.

The NIHAO simulations form the right amount of stars and cold gas as
evidenced by  consistency with the stellar mass vs halo mass relations
from abundance matching since $z=4$ \citep{Wang15}, and the cold gas
vs stellar mass relation at $z=0$ \citep{Stinson15}.  In
Fig.~\ref{fig:msmh} we show the stellar mass vs halo mass relation for
the mass scale relevant to this study and compare to the latest
abundance matching results  \citep{GK14a,Brook14}, which are
consistent with the Local Group stellar mass function. Here the
  relation from \citet{Brook14} is normalized assuming a Milky Way
  halo mass of $1.4\times 10^{12}\Msun$ \citep{Brook15}. The dotted
  lines correspond to where the relations are extrapolated.  The
shaded region shows the 1$\sigma$ scatter of 0.23 dex determined by
\citet{Behroozi13}.  Our stellar mass is measured within 20 percent of
the virial radius. In contrast to other NIHAO papers, here we define
the halo mass with an overdensity of 100 $\times$ the critical
density, to be consistent with \citet{GK14a}.

The agreement between our simulations and abundance matching implies
that sampling the  halo mass function would reproduce the number
densities of observed galaxies. Thus rather than explicitly counting
the number of massive failures, as in \citep{GK14b}, we require
our simulations reproduce the average scaling relations between
circular velocity and size.

\section{Scaling relations}
Fig.~\ref{fig:vlr} shows the size-luminosity and velocity-luminosity
relations for NIHAO simulations with luminosities $ 0.5\times 10^5
< \LV/\Lsun < 2\times 10^8$ (red filled symbols) vs observations (open
symbols) from \citet{Kirby14}. In the simulations we calculate 3D
half-light radii, $r_{1/2}$, using the cumulative V-band luminosity
profile \citep[computed with {\sc pynbody,}][]{Pontzen13} inside 20\%
of the virial radius.  Note that since the sizes are typically 2
  per cent of the virial radius, they are insensitive to the exact
  choice of outer aperture. We measure the circular velocity at the
half-light radius, $V_{1/2}=\sqrt{G M(r_{1/2})/r_{1/2})}$.  We convert
the observed 2D half-light radii into 3D half-light radii by
multiplying by 4/3. The observed circular velocities at the half-light
radius are computed by converting the $M_{1/2}$ from \citet{Kirby14}
into $V_{1/2}$. In the absence of rotation, this corresponds to
$V_{1/2}=\sqrt{3}\sigma$. 

The simulations correspond well with observations,  especially for the
isolated galaxies (black open symbols) which are a fairer comparison
sample to our isolated simulated galaxies.  Below $L_V \sim 2\times
10^6 \Lsun$ there is a large observational scatter in both relations,
with no clear trend, while above $L_V \sim 2\times 10^6 \Lsun$ the
scatter is smaller and there are clear trends for more luminous
galaxies to be larger with higher circular velocities. We note that at
least part of the increased scatter at low luminosities is
observational, and especially including M31 galaxies.

The fact that our simulations reproduce the stellar mass-halo mass,
size-luminosity, and velocity-luminosity relations would suggest that
they also resolve the too-big-to-fail problem.

\begin{figure*}
  \centerline{
    \includegraphics[width=0.94\textwidth]{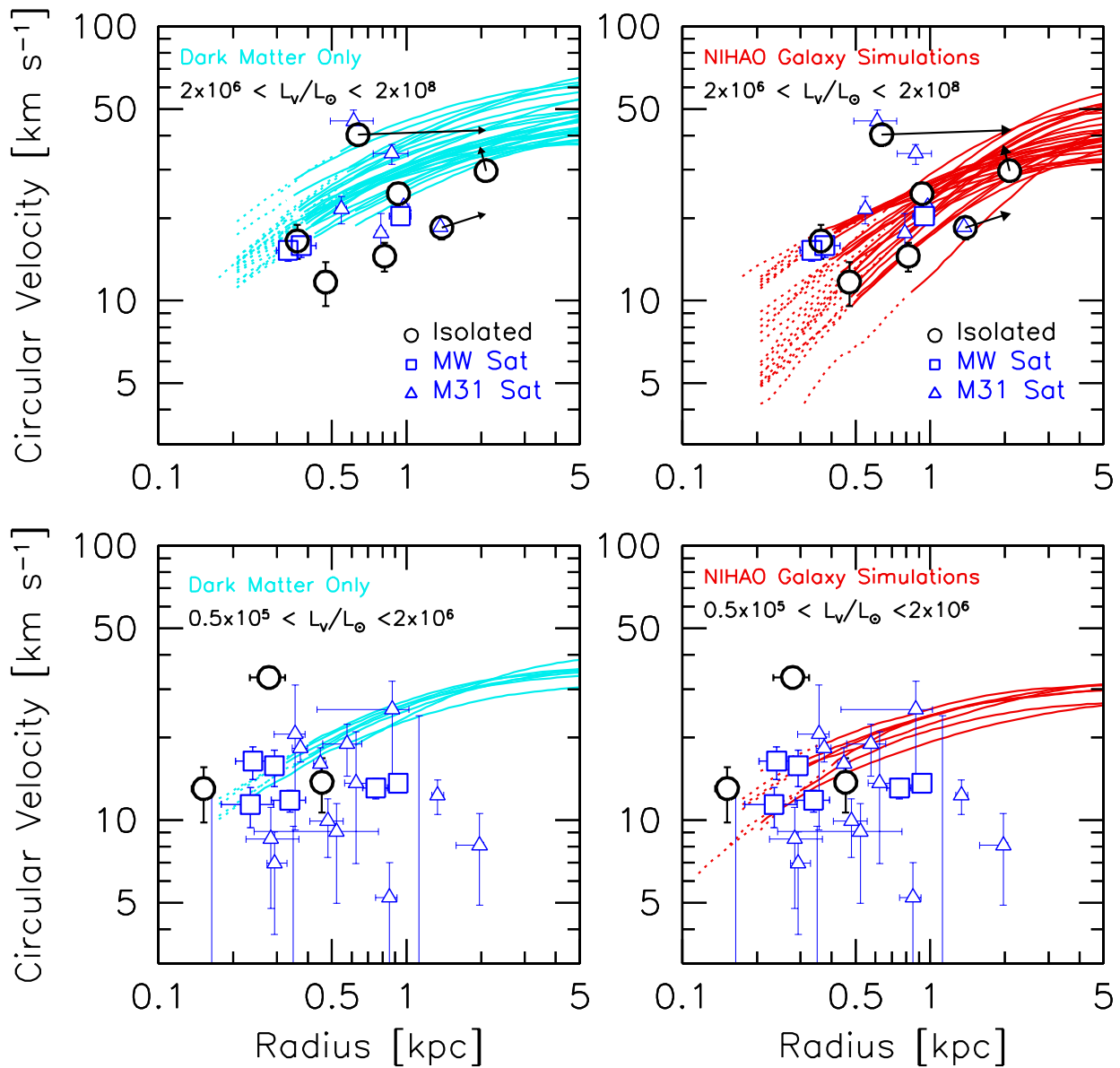}
}
\caption{Circular velocity vs radius for cosmological simulations
  (lines) compared to observations (symbols with error bars) of local
  group galaxies from Kirby \etal (2014), split into isolated (black)
  and satellites of the Milky Way (blue squares) and M31 (blue
  triangles). Upper and lower panels show galaxies greater and  less
  than  a luminosity of $L_V = 2\times 10^6 \Lsun$, respectively.
  The arrows for the three brightest isolated galaxies show
  alternative measurements of circular velocity at 2 kpc.  The left
  panels show dark matter only simulations (in cyan), while the right
  panels show the NIHAO galaxy simulations (in red). The solid lines
  show the simulated profiles down to where the velocity profile has
  converged to 10\%, while the dotted lines continue the profile to
  the dark matter softening length.} 
    \label{fig:vr}
\end{figure*}

\section{Too big too fail}
Fig.~\ref{fig:vr} shows the circular velocity profiles of our
simulations (with luminosities $ 0.5\times10^5 < \LV/\Lsun <
  2\times 10^8$) vs observations of satellite and isolated galaxies
in the local group compiled by \citet{Kirby14}. Simulations are
plotted with solid lines down to the radius where the circular
velocity profile has converged to 10\% according to the criteria of
\citet{Schaller15}. The dotted lines continue the profiles down to the
softening length of the dark matter particles. 

It has been noted that isolated galaxies (black circles) fall in the
same part of the velocity vs radius plane as the satellites of the
Milky Way (blue squares) and M31 (blue triangles). This suggests that
environmental processes do not significantly affect the average dark
halo mass profile, limiting the range of physical processes
available to solve the TBTF problem in the context of $\LCDM$.

We split the sample into two luminosity groups at $L_V=2\times 10^6\Lsun$
as galaxies above and below this luminosity appear to have
qualitatively different behavior. Observationally, the low luminosity
galaxies have large variation in sizes and velocities at fixed
luminosity, while the high luminosity galaxies follow well defined
scaling relations (given the small sample sizes).  Theoretically, at
low luminosities the stellar mass vs halo mass relation breaks down,
and the halo response to galaxy formation is minimal.

The cyan lines in the left-hand panels of Fig.~\ref{fig:vr} show the
circular velocity profiles of the dark matter only simulations. At
both high and low luminosities these simulations predict
systematically higher velocities than observed.  For low-luminosity
galaxies the median half-light size and circular velocity of observed
galaxies is $r_{1/2}=0.37$ kpc and $V_{1/2}=12.3\kms$. At 0.37 kpc the
median circular velocity in the dark matter only simulations is
$16.1\kms$.  For high-luminosity galaxies the median half-light size
and circular velocity of observed galaxies is $r_{1/2}=0.79$ kpc and
$V_{1/2}=18.6\kms$. The corresponding median circular velocity
in the dark matter only simulations is $28.9\kms$, i.e. 55\% too high.
This discrepancy in circular velocity corresponds to a factor of $\sim
2$ in enclosed mass, and is consistent with previous studies of the
TBTF problem of satellite and field galaxies \citep{BK12,GK14b}.

\section{A baryonic solution to TBTF}

In the right-hand panels of Fig.~\ref{fig:vr} the red lines show the
circular velocity profiles of the NIHAO galaxy simulations. Focusing
first on the upper panel, there is no systematic offset and the
scatter is comparable, showing that the effects of galaxy formation,
and particular the prescription for star formation and stellar
feedback implemented in NIHAO cause the right amount of halo expansion
(a factor of 1.5 lower circular velocity at 0.79 kpc).  An interesting
feature of the halo response is that the more massive haloes have
larger reductions in central rotation velocities, and shallower
  density slopes \citep{Tollet15}. This causes the rank order of
circular velocity at sub-kpc scales to no-longer correspond to the
rank order of their halo masses. 

For the three most luminous isolated galaxies (NGC6822, IC1613, WLM)
the arrows show a measurement of the circular velocity at 2 kpc from
resolved neutral hydrogen rotation curves as compiled by
\citet{Oman15}.  With these observations NGC6822 is now consistent
with the hydro simulations, and our galaxy formation simulations are
consistent with all of the luminous field galaxies.

\section{A problem for low mass galaxies?}
Lower luminosities and halo masses are potentially interesting, as our
(and other) simulations predict that baryonic processes have
negligible impact on the structure of dark matter haloes. For
luminosities below $L_V\sim 2\times 10^6\Lsun$ the halo response is
minimal: the median circular velocities at 0.37 kpc in the NIHAO
simulations are just 2\% lower than the dark matter only simulations.

There are only three observed isolated galaxies in this luminosity
range, two are consistent with our simulations, while one (Tucana --
black circle at $r_{1/2}\simeq 0.3$ kpc, $V_{1/2}\simeq 30 \kms$) is
significantly above. If the measurements of the circular velocity and
half-light size are robust, then it must have formed  in a massive
dark matter halo with $V_{\rm max} \gta 70 \kms$.  Some of the
satellites are consistent with the simulations, but half are
significantly below, including two MW satellites (although it should
be noted that most of the discrepant galaxies are M31 satellites and
carry larger measurement errors).

The larger scatter in observed circular velocities than predicted by
our simulations may indicate a lingering problem for $\LCDM$.  This
problem is distinct from the TBTF problem, as the typical host haloes
are no longer too-big-to-fail to form stars.  A possible solution is
that there is a wide range of halo masses that host galaxies of
luminosity $ 10^5 \lta L_V/\Lsun \lta 10^6$. Specifically, if some
haloes of mass $\Mhalo \lta 10^9 \Msun$ are efficient at forming stars
($\Mstar \gta 10^5 \Msun$), then the lower part of the velocity
radius plane could be filled up.  Indeed in our own simulations we see
the tight stellar mass vs halo mass relation breaks down in haloes
below $\Mhalo \sim 10^{10}\Msun$, see also \citet{Sawala15}.

Larger samples of high-resolution simulated galaxies (both field and
satellite) are clearly needed to fully sample the stochasticity of
galaxy formation in haloes below $\Mhalo \sim 10^{10}\Msun$. More
observations of low luminosity field galaxies would help to clarify
the observational picture.

\section*{Acknowledgements}

We thank the referee for a prompt and constructive report.
The simulations were performed on the {\sc theo} cluster of the
Max-Planck-Institut f\"ur Astronomie and the {\sc hydra} cluster at the
Rechenzentrum in Garching; and the Milky Way supercomputer, funded by
the Deutsche Forschungsgemeinschaft (DFG) through Collaborative
Research Center (SFB 881) ``The Milky Way System'' (subproject Z2),
hosted and co-funded by the J\"ulich Supercomputing Center (JSC). We
greatly appreciate the contributions of all these computing
allocations.
AAD, AVM, and GSS acknowledge support through the
Sonderforschungsbereich SFB 881 ``The Milky Way System'' (subprojects
A1 and A2) of the German Research Foundation (DFG).
JF acknowledges funding through the graduate college {\em Astrophysics of
cosmological probes of gravity} by Landesgraduiertenakademie
Baden-W{\"u}rttemberg.
LW acknowledges support of the MPG-CAS student programme.
GSS acknowledges funding  from the European Research Council under the
European Union's  Seventh Framework Programme (FP 7) ERC Grant
Agreement n. [321035].
XK acknowledge the support from NSFC project
No.11333008 and the "Strategic Priority Research Program the Emergence
of Cosmological Structures" of the CAS(No.XD09010000).







\bsp	
\label{lastpage}
\end{document}